\documentclass[%
preprint,
superscriptaddress,
amsmath,amssymb,
aps,
pra,
]{revtex4-2}

\usepackage{graphicx}
\usepackage{dcolumn}
\usepackage{bm}
\usepackage{braket}
\usepackage{amsmath,amssymb,amsfonts}
\usepackage{mathrsfs}
\usepackage[colorlinks,linkcolor=blue,citecolor=blue,urlcolor=blue]{hyperref}

\renewcommand\a{\alpha}
\renewcommand\b{\beta}

\newcommand\del{\delta}

\renewcommand\r{\rho}

\renewcommand\j{\psi}

\newcommand\G{\Gamma}

\newcommand\J{\Psi}

\def\ket#1{\lvert#1\rangle}

\newcommand{\no}{\nonumber}

\begin{document}
	
	
	\title{Coherent States in Classical Field Theory}
	
	\author{Abhijeet Joshi}
	\author{Vivek M. Vyas}%
	\affiliation{%
		Indian Institute Of Information Technology Vadodara
		\\
		Government Engineering College, Sector 28, Gandhinagar
	382028, India}%
	\author{Prasanta K. Panigrahi}%
	\affiliation{%
		Center for Quantum Science and Technology, Siksha `O' Anusandhan University,\\ Bhubaneswar, Odisha 751030, India
		}
	\affiliation{Department of Physical Sciences, Indian Institute of Science Education \& Research (IISER) Kolkata,\\ Mohanpur, Nadia 741 246, India}
	
	\date{\today}
	
\begin{abstract}
We illustrate the emergence of classical analogue of coherent state and its generalisation in a purely classical field theoretical setting. Our algebraic approach makes use of the Poisson bracket and symmetries of the underlying field theory, in a complete parallel to the quantum construction. The classical phase space picture is found to play a key role in this construction. 
\end{abstract}
	
	\maketitle
	
\section{\label{intro} Introduction}

In the literature there exists several notions of the concept of coherent state \cite{RevModPhys.62.867,glauber1963quantum,sudarshan1963,klauder1985coherent,perelomov1977generalized,roy1982generalized,shanta1994unified,twareque1995coherent,agarwal2012quantum}. In the celebrated quantum harmonic oscillator, it is well known that the  existence of Heisenberg-Weyl algebra:
\begin{align}
	&[\hat{a},\hat{a}^{\dagger}] = \mathbf{1}, \: \text{where} \\
     &\hat{a} = \frac{(\hat{x} + i \hat{p})}{\sqrt{2}},
\end{align}
is responsible for the genesis of the coherent states. These coherent states are known to obey the following properties:

\begin{itemize}
	\item[A.] They are eigenstates of annihilation operator $\hat{a}$: $\hat{a} \ket{\a} = \a \ket{\a}$,
	
	\item[B.] They are obtained by the action of the group element $\hat{\mathscr{D}}(\a) = e^{\a a^{\dagger} - \a^{\ast} a}$ on the ground state (defined as $\hat{a}\ket{0} = \ket{\emptyset}$): $\ket{\a} = \hat{\mathscr{D}}(\a) \ket{0}$, 

    \item[C.] They saturate the Heisenberg uncertainty relation, in the sense that they are minimum uncertainty states,
    
    \item[D.] Their shape does not alter under time evolution, in the sense that their wavepacket does not spread or disperse with time, and

    \item[E.] The center or the peak of the wavepacket follows the classical equation of motion.
\end{itemize}
Over the years various generalisations of the notion of coherent state in various other physical system have been proposed and studied. The well known generalisation due to Perelomov \cite{perelomov1977generalized}, relies on the Lie group structure in the problem at hand, essentially generalising the property (B) \cite{RevModPhys.62.867,twareque1995coherent}. While the work of Singh and Roy \cite{roy1982generalized} considers the properties (D) and (E) to be essential for a wavepacket to be a coherent state. In a recent work Englert \cite{englert2024uncertainty} has studied various minimum uncertainty wavepackets. There have also been several in-depth studies on the various mathematical and physical aspects of these generalisations \cite{RevModPhys.62.867,klauder1985coherent,twareque1995coherent, puri2001mathematical}. Coherent states are often described as the bridge between the quantum and classical descriptions of a physical system.

In the present work, we focus our attention to the work of Shanta \emph{et. al.,} \cite{shanta1994unified} wherein a large class of coherent states were found to essentially solve the eigenvalue problem of certain operators in the theory. This work in essence relies upon generalisation of property (A). Here we show that several well known coherent states admit a purely classical field theoretic counterpart in a one-dimensional nonrelativistic complex scalar theory. We note the key role played by the notion of position variable $X$, which is constructed so as to be the canonical conjugate of the momentum $P$. The existence of the phase space structure in the classical theory is the key factor of this construction, which allows us to translate the concepts and ideas from quantum Hilbert space into a completely classical setup. 

\section{Classical Field Theory\label{sec2}}

In this work we shall be considering a one-dimensional nonrelativistic complex scalar field theory of field $\psi(x,t)$. Let the Lagrangian density be given by:
\begin{equation} \label{action}
	\mathscr{L} = \psi^{*}\Bigl(i \r \frac{\partial\psi}{\partial t}+
	\b\frac{ \partial^2 \psi}{\partial x ^2}\Bigr) + \mathscr{L}_{int}(\psi^*,\psi) 
\end{equation} 
where $\r$ and $\b$ are some constants. $\mathscr{L}_{int}$ stands for the possible local self-interaction terms of the form $g|\psi|^4$ for example, giving rise to nonlinearity. { In this discussion, we shall confine our attention to the set S of square integrable field configurations which decay sufficiently fast to spatial infinity: $\psi(x,t) \rightarrow 0$ as $|x| \rightarrow \infty$, so that the integral $\int dx |\psi(x,t)|^2$ is finite. The rationale behind considering this boundary condition essentially comes from the fact that in the quantised field theory $\int dx |\psi(x,t)|^2$ denotes the total particle number, which for a large class of physical systems is finite.}

The canonically conjugate momentum $\psi(x,t)$ is given by:
\begin{equation}
	\pi(x,t) = i \r \psi^{\ast}(x,t) \label{conjmom},
\end{equation}
which allows us to define the equal time Poisson bracket between any two quantities $A$ and $B$ as \cite{greiner2013field}:
\begin{align} \no
	&\{A, B\} \\
	&= \int \: dz \: \left( \frac{\del A}{\del \j(z,t)} \frac{\del B}{\del \pi(z,t)} - \frac{\del B}{\del \j(z,t)} \frac{\del A}{\del \pi(z,t)} \right). \label{poissonb}
\end{align}
The canonical equal time Poisson bracket in this theory reads:
\begin{align} \label{canonicalpb}
	\{ \j(x,t) , \pi(y,t)  \} = \del(x-y).
\end{align}

{ Invoking Noether theorem \cite{greiner2013field,das2020lectures} one sees that this theory, whose action is invariant under continuous spatial translation, admits momentum $P$ as the generator of spatial translations:
\begin{equation} \label{momentum}
	P=\int dx \psi^*(x,t)\frac{\r}{i}\frac{\partial\psi}{\partial x},
\end{equation}
so that we have:
\begin{align}
	 \{P,\psi \} = \frac{\partial \psi }{\partial x}. \label{gentrans}
\end{align}
}
Let us define position variable $X$ in the theory as:
\begin{equation} \label{position}
	X=  \int dx \psi^*(x,t) x \psi(x,t).
\end{equation}
For here it immediately follows that its equal time Poisson bracket with $P$ is:
\begin{align} \no
	\{X,P\}&= \int dy \psi^*(y) \psi(y)\\ &= N, \label{canonicalpb}
\end{align}
where $N = \int dx \psi^*(x) \psi(x)$. { So long as $N$ is finite one sees that $X$ is canonically conjugate to $P$, under the equal time Poisson bracket (\ref{poissonb}). It is worth noting that the Poisson bracket $\{X,\psi\}$ is: 
\begin{align}
	\{{X,\psi}\} &= \frac{i}{\r} x \psi. 
\end{align}
}

{Invoking Noether theorem it is easy to see that  $N = \int dx \psi^*(x) \psi(x)$ is the generator of global phase transformations: $\psi(x,t) \rightarrow \psi(x,t) e^{i \theta}$ \cite{das2020lectures}. As noted earlier, $N$ in the case of quantised field theory denotes the total particle number, as also the electric charge in the theory \cite{das2020lectures,greiner2013field}. In most physical scenarios, the total particle number and electric charge are conserved under time evolution, and so is $N$. We shall henceforth assume this to be the case in what follows. It must be kept in mind that the only restriction this puts on the Lagrangian (\ref{action}) is that it must be invariant under the global phase transformation $\psi(x,t) \rightarrow \psi(x,t) e^{i \theta}$. The nonlinear self interaction terms of the type $\mathscr{L}_{int} = g|\psi|^n$ are simple yet nontrivial terms that admit such an invariance.}

\section{Coherent states}

In the earlier section, we saw the appearance of the canonically conjugate pair of position $X$ and momentum variable $P$ in the classical field theory, defined by action (\ref{action}). This construction motivates us to define the classical analogue of annihilation operator $\hat{a}$ as $a_{c}$:
\begin{equation}
a_{c} = \frac{ X + i P}{\sqrt{2 N}}.	
\end{equation}
The equal time Poisson bracket $\{a_c,a_c^*\}$ can be straight away calculated to read: 
\begin{align} \label{classalg}
\{a_c,a_c^*\}=-i.	
\end{align}
One can further construct the analogue of number operator  $\hat{n}=\hat{a}^\dagger\hat{a}$ of harmonic oscillator as: $N_c = a_c^* a_c$ so that the classical version of harmonic oscillator algebra holds:
\begin{align}
\{N_c,a_c\} = i a_c, \:	\{N_c,a_c^*\}  =-ia_c^*.	
\end{align}

This construction makes us capable of now asking the following question: Coherent states in quantum harmonic oscillator are defined as eigenstates of annihilation operator $\hat{a} \ket{\alpha} = \alpha \ket{\alpha}$, is it possible to define classical field theoretic counterpart coherent states $\ket{\a}$ using $a_{c}$ ?
\par
A crucial question that one has to answer first is that what is the classical counterpart of the eigenvalue equation $\hat{a} \ket{\alpha} = \alpha \ket{\alpha}$ ? 

Here we propose that it should be $\{a_c,\J_{\alpha}(x)\} = i \alpha \J_{\alpha}(x)$, where $\J_{\alpha}(x)$ represents a particular field configuration belonging to S. 

Thus the classical analogue of $\ket{\alpha}$ can be easily obtained by solving this equation to yield:
\begin{equation}
	\J_{\alpha}(x)= \text{constant} \times e^{- \frac{\r}{2}x^2} e^{\sqrt{2N}\a x}.
\end{equation}
which is identical to the position representation of coherent state $\ket{\a}$ in quantum mechanics \cite{agarwal2012quantum}.

This motivates us further to find the classical analogue of cat states, which are the eigenstates of $\hat{a}^2$ \cite{shanta1994unified}, as:
\begin{gather}
	\{a_c,\{a_c,\Phi_{\alpha}(x)\} \} =\alpha^2\Phi_{\alpha}(x).
\end{gather}
This can be readily solved to yield $\Phi_{\alpha}(x)=\J_{-\alpha}(x)+\J_{\alpha}(x)$.

Further more, we can go on and also construct a classical analogue of kitten or compass state, which are eigenstates of $\hat{a}^4$ \cite{shanta1994unified,zurek2001sub}, as:
\begin{equation}
	\{a_c,\{a_c,\{a_c\,\{ a_c,\G(x) \} \} \} \}=\alpha^4 \G(x),
\end{equation}
which yields $\G(x)=\J_\alpha (x)+\J_{-\alpha} (x)+\J_{i\alpha}(x)+\J_{-i\alpha}(x)$.

In fact one can construct the classical counterpart of what is called Agrawal-Tara state or photon added coherent state \cite{agarwaltara}  $ \hat{a}^{\dagger} \ket{\a}$ as:
\begin{align}\no
	Y_{1,\a}(x) &= \{ a_{c}^{*} , \J_{\alpha} (x) \} \\
				&= i \left( \sqrt{\frac{2}{N}} \r x - \a \right) \J_\alpha (x).		
\end{align}

This discussion might make the reader wonder if only the coherent states that owe their genesis to quantum harmonic oscillator admit classical counterparts. This is not the case, as we show below that well known self accelerating Airy wavepackets, which are known to be an example of Perelomov coherent states \cite{vyas2018airy}, also admit a classical counterpart. 

Let us consider the case of free field theory whose  Hamiltonian is given by:
\begin{equation}
	H=-\int dx \b \psi^* \frac{\partial^2\psi}{\partial x^2}, 
\end{equation}
and the generator of Galilean boosts is $K=tP-NX$. From \citep{vyas2018airy} it is known that Airy wavepacket solve the eigenvalue problem for $K+\epsilon H$ ($\epsilon$ is a real parameter). This motivates us to seek the field configuration $\chi(x)$ that solve: 
\begin{gather}
	\{K+\epsilon H,\chi \} =iz \chi,
\end{gather}
where $\epsilon$ and $z$ are real parameters. This results in the differential equation:
\begin{gather}
	t \frac{\partial \chi}{\partial x}-\frac{iNx}{\r}\chi - i \epsilon \b \frac{\partial ^2\chi }{\partial x ^2}-iz \chi=0,
\end{gather}
whose solutions are self-accelerating and non-dispersing Airy functions:
\begin{align} \no
	\chi(x,t)&= e^{-\frac{itx}{2\b \epsilon}} \left( c_{1} \text{Ai} \left[ -(- \frac{\r \b \epsilon}{N})^{\frac{2}{3}} (  \frac{Nx}{\r \b \epsilon} + \frac{t^2}{4 \b^2 \epsilon^2} + \frac{z}{\b \epsilon}  ) \right]
    \right. \\& \left. + c_{2} \text{Bi} \left[ -(- \frac{\r \b \epsilon}{N})^{\frac{2}{3}} (  \frac{Nx}{\r \b \epsilon} + \frac{t^2}{4 \b^2 \epsilon^2} + \frac{z}{\b \epsilon}  ) \right] 
	\right),
\end{align}
where $c_1$ and $c_2$ are real constants. 

\section{Conclusion}

In this work we have shown that various well known coherent states, that are defined completely in a quantum mechanical framework, admit a classical counterpart in a one-dimensional complex nonrelativistic scalar theory. The key step in this construction is the definition of a position variable $X$ in the field theory that is canonically conjugate to momentum $P$. Furthermore the existence of canonical Poisson bracket structure in the classical field theory is found to play pivotal role as it provides a natural algebraic framework to translate the ideas and notions defined over the quantum Hilbert space to the classical phase space. Thus we have found the classical counterparts of harmonic oscillator coherent states, cat state, kitten or compass state, self accelerating Airy coherent state, and photon added coherent state.


%

\end{document}